\documentclass[pre,twocolumn,showpacs,floatfix]{revtex4-1}
\usepackage{graphics,epsfig,amsfonts,amssymb,amsmath,soul}
\usepackage[T1]{fontenc}
\usepackage[utf8]{inputenc}
\usepackage{lmodern}
\usepackage[english]{babel}
\usepackage{wasysym}
\usepackage{color}
\begin{document}
\selectlanguage{english}
\title{Moving magnets in a micromagnetic finite difference framework}

\author{Ilari Rissanen}
\email{ilari.rissanen@aalto.fi}
\author{Lasse Laurson}

\affiliation{COMP Centre of Excellence and Helsinki Institute of Physics,
Department of Applied Physics, Aalto University, P.O.Box 11100, 
FI-00076 Aalto, Espoo, Finland.}

\begin{abstract}
We present a method and an implementation for smooth linear motion in a finite difference-based micromagnetic simulation code, to be used in simulating magnetic friction and other phenomena involving moving microscale magnets. Our aim is to accurately simulate the magnetization dynamics and relative motion of magnets while retaining high computational speed. To this end, we combine techniques for fast scalar potential calculation and cubic b-spline interpolation, parallelizing them on a Graphics Processing Unit (GPU). The implementation also includes the possibility of explicitly simulating eddy currents in the case of conducting magnets. We test our implementation by providing numerical examples of stick-slip motion of thin films pulled by a spring and the effect of eddy currents on the switching time of magnetic nanocubes.
\end{abstract}
\maketitle

\section{Introduction}

Numerical micromagnetics is often the tool of choice when investigating the behavior of magnetization at scales where fine details of magnetic structures such as domain walls need to be resolved but atomic description is computationally unfeasible. Since their early application in predicting domain wall structure in soft thin films \cite{micromag1}, micromagnetic simulations have been used to reproduce a variety of experimental results  \cite{microexp1,microexp2, microexp3}. With the advances in GPU-accelerated computing, the speed of micromagnetic simulations has surged  \cite{gpu1,mumax1}, making it possible to perform larger length- and timescale simulations.

An area currently lacking in micromagnetics is the capability of simulating mechanical motion of magnets and the interplay of motion and the domain dynamics of magnets moving relative to each other. The relative motion of small-scale magnets is relevant in studying phenomena such as magnetic friction  \cite{AtomicSpinFriction, IsingFric, motioncontrol, magiera1, magiera4}, and for applications such as magnetic force microscopy  \cite{MFM}, and micro- and nanomanipulation \cite{manipul4, manipul3}. Despite the scientific interest in these areas, simulation frameworks capable of general micromagnetic simulations coupled with the motion dynamics of the magnets, to the authors' knowledge, do not exist. In magnetic friction context, studies have been performed on specific simulation instances such as single magnetic dipole being moved at a constant velocity atop a monolayer  \cite{magiera1, magiera2} and perpendicularly polarized thin films sliding relative to each other  \cite{motioncontrol}, but otherwise computational studies of moving microscale magnets and their interactions have been scarce. 

In this paper, we extend an existing finite difference micromagnetic simulation code to handle the linear motion of a magnet interacting with another. In the case of conducting magnets, we also include an eddy current solver in our movement implementation in order to study the effects of eddy currents on the motion and magnetization dynamics. Our primary focus is on magnetic friction and thin films, but other phenomena involving moving magnets can also be studied within the framework. 

The structure of the paper is as follows: in Section~\ref{sec:movement}, we examine the simulation of moving magnets in a finite difference framework and present our method for simulating smooth motion. Section \ref{sec:movImplementation} elaborates on the technical details of the movement and eddy current implementations. In Section \ref{sec:results}, we test our implementation with example simulations, comparing the obtained results to those of previous works on magnetic friction \cite{motioncontrol} and eddy currents \cite{TorresEddy, eddycurrenthead}. Finally, in Section \ref{sec:summary} we summarize the main points of the article and conclude with thoughts on possible future work.

\section{Moving microscopic magnets}
\label{sec:movement}

At the core of micromagnetics is the the Landau-Lifshitz-Gilbert (LLG) equation, which governs the time evolution of the magnetization in a magnetic material. It can be written as
\begin{equation}
\frac{\partial \textbf{m}}{\partial t}=-\gamma \mathbf{H}_{\mathrm{eff}} \times \mathbf{m} + 
\alpha \mathbf{m} \times \frac{\partial \textbf{m}}{\partial t},
\end{equation}
where $\gamma$ is the gyromagnetic ratio, $\mathbf{m}$ is the normalized magnetization $\mathbf{m}=\mathbf{M}/M_\mathrm{sat}$,  $\alpha$ is the phenomenological Gilbert damping constant, and $\mathbf{H}_\mathrm{eff}$ the effective field, in most cases containing four field terms: exchange field $\mathbf{H}_\mathrm{exch}$, anisotropy field $\mathbf{H}_\mathrm{anis}$, external (Zeeman) field $\mathbf{H}_\mathrm{ext}$ and demagnetizing field $\mathbf{H}_\mathrm{d}$. In micromagnetics, the magnetic properties of a material are determined by material parameters such as the exchange constant $A_\mathrm{ex}$, the saturation magnetization $M_\mathrm{sat}$ and the Gilbert damping constant.

In numerical micromagnetics, finite difference methods have been found attractive due to the possibility of using Fast Fourier transforms to speed up the evaluation of the demagnetizing field, which usually is the computationally most demanding part of micromagnetic simulations  \cite{numericalmicromagnetics}. In finite difference micromagnetics, the domain of interest is discretized into cuboid (often cubic) cells in which the LLG equation is solved. A discretization cell can be empty (nonmagnetic, vacuum/air) or contain magnetic material, in which case the cell is typically treated as uniformly magnetized, represented by a vector in the center of the discretization cell. These magnetization vectors interact with the local field $\mathbf{H}_\mathrm{eff}$ in each cell and evolve in time according to the LLG equation. 

\begin{figure}[t!]
\leavevmode
\includegraphics[trim=0cm 0cm 1.0cm 0.2cm, clip=true,width=1.00\columnwidth]{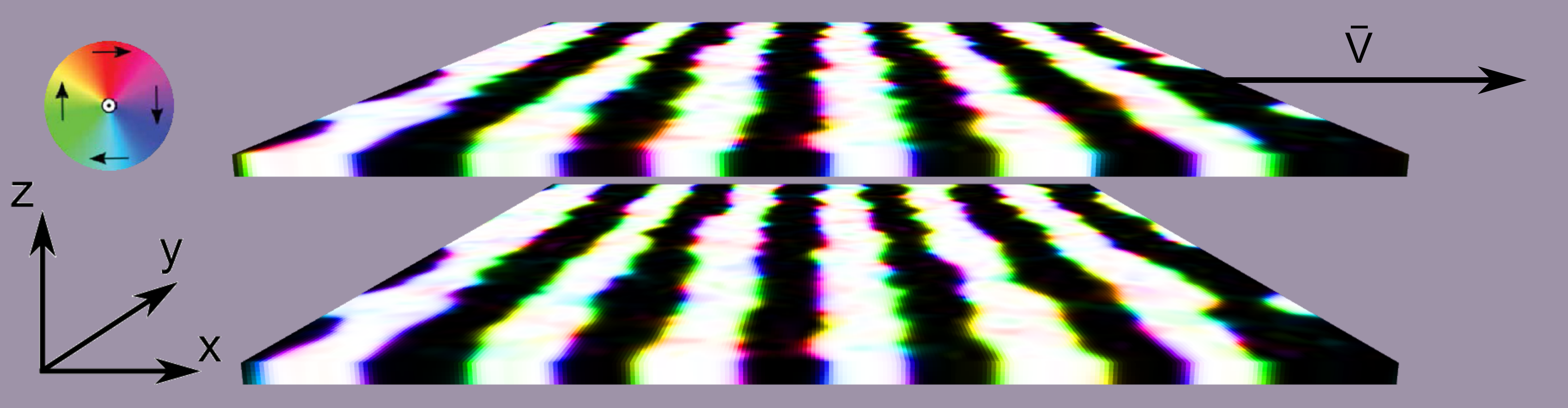}
\caption{An example system of two stripe-patterned thin films for simulating magnetic friction. In this example, the upper film is driven towards $+x$-direction with a constant velocity $v$ and the system is periodic in the film plane (periodic images not shown). The direction of magnetization in the $xy$-plane is indicated by the color wheel on the left, while white and black correspond to magnetization in $+z$ and $-z$ directions, respectively.}
\label{FIGSimulationExample}
\end{figure}

\begin{figure}[t!]
\leavevmode
\includegraphics[trim=0cm 0cm 0cm 0cm, clip=true,width=1.00\columnwidth]{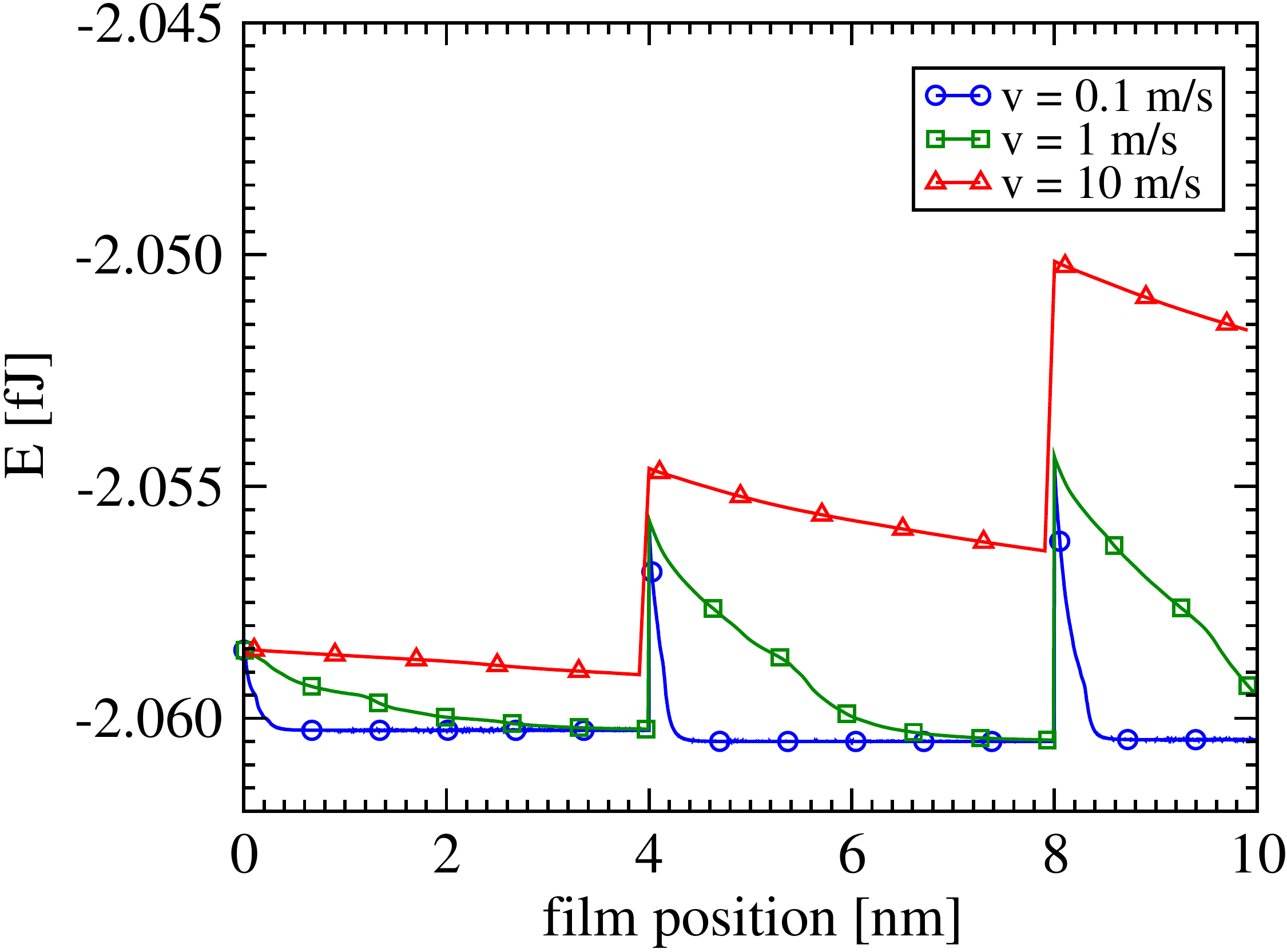}
\caption{A demonstration of the discontinuous jumps in total energy caused by the discrete movement in the example system depicted in Fig.~\ref{FIGSimulationExample} for three different driving velocities.}
\label{FIGEnergyJumps}
\end{figure}

A possible simulation scenario with two magnetic thin films in relative motion is depicted in Fig.~\ref{FIGSimulationExample}. Movement in this kind of simulation scheme consists of updating the cells with the correct material parameters and magnetization vectors when the magnet moves in and out of the simulation cells. However, due to the discretization the magnet can only move in discrete jumps inside the simulation domain, and thus $\mathbf{m}$, $\mathbf{H}_\mathrm{eff}$ and other quantities inside the cells change discontinuously in response to the motion. This leads into a cycle of excitations and relaxations of the magnetization, which is especially prominent at low velocities. The intermittent movement also introduces artificial discontinuities in quantities such as energy of the system (Fig.~\ref{FIGEnergyJumps}), complicating the analysis of results  e.g. when studying stick-slip motion. Additionally, when eddy currents are included in the simulation (see Section~\ref{sec:EddyImplementation}), the discontinuous change in the local magnetic field due to the jumps could lead to an overestimation in the induced eddy currents. In order to properly study the interaction of two magnets in relative motion and the related phenomena, smooth continuous motion is desirable.

\subsection{Smooth motion through interpolation}
There are a couple of ways to realize smooth motion within a finite difference framework. The simplest idea is to use smaller simulation cells to limit the size of the cell-to-cell jumps during the motion, but this increases the computational cost dramatically and does not truly eliminate the problem. Another way is to emulate movement between cells by scaling the $M_\mathrm{sat}$ of the partially filled cells by the percentage of the cell volume containing magnetic material, but this can lead to errors unless corrections are made to the calculation of  exchange interaction and demagnetizing field terms  \cite{Garcia-Cervera}. When simulating magnets in motion, applying the corrections could become computationally quite intensive. A simpler approach applicable to our simulation scenario, that is two magnets in relative motion and not in direct contact, is to use interpolation to find the effective field in between the discretization cells. This approach can be made computationally quite inexpensive when appropriate calculation methods are used.

\begin{figure}[t!]
\leavevmode
\includegraphics[trim=0cm 0cm 0cm 0cm, clip=true,width=0.93\columnwidth]{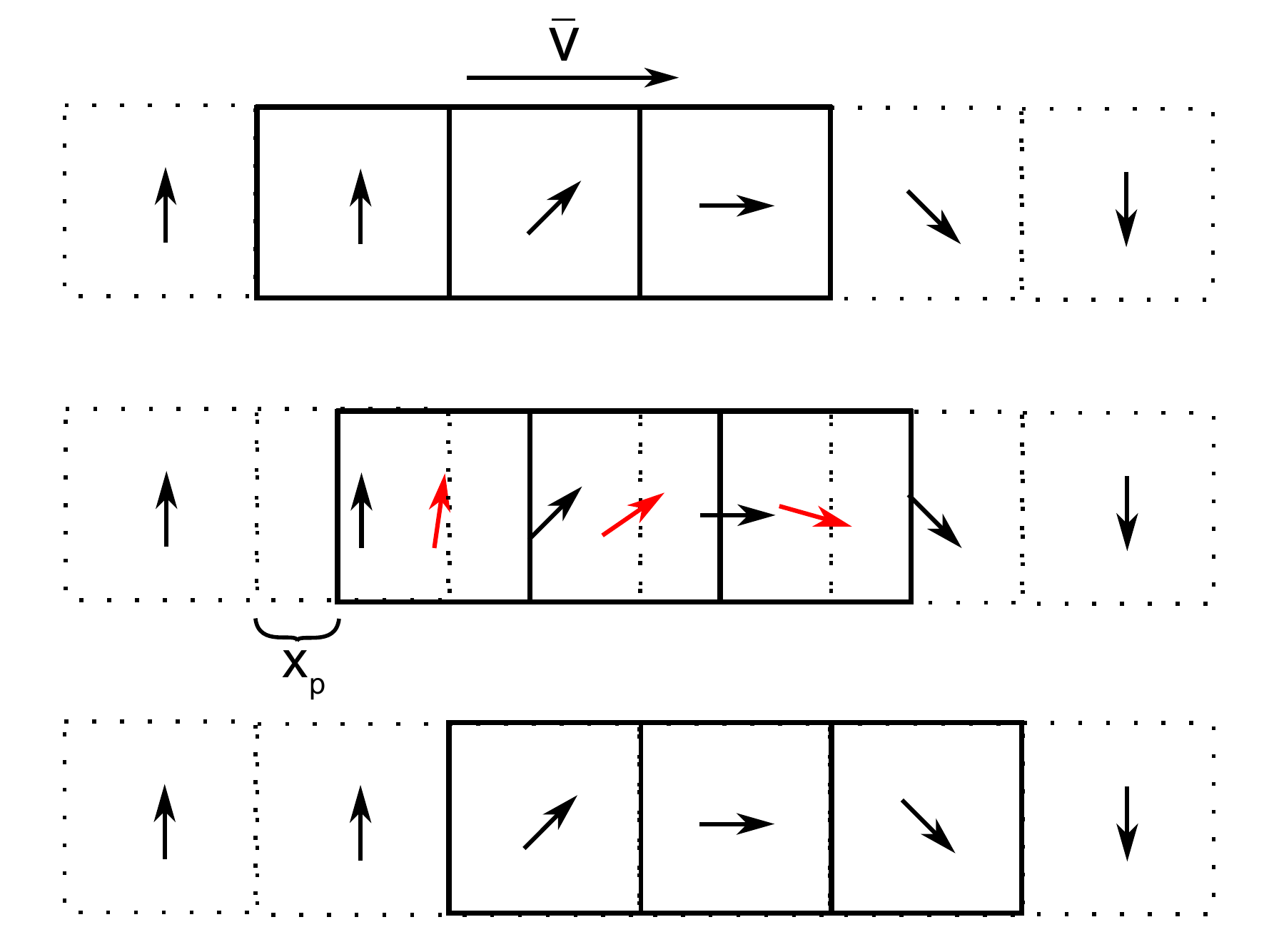}
\caption{An illustration of a part of a discretized magnet moving in between the simulation cells with speed $\mathbf{v}$ in the $x$-direction, the partial movement denoted by $x_\mathrm{p}$, the black arrows depicting the demagnetizing field vectors of the stationary magnet and the interpolated field vectors shown in red inside the moving magnet.}
\label{FIGInterpolation}
\end{figure}

Two magnets that are not in direct contact interact with each other only via the long range interaction term of the LLG equation, the demagnetizing field (or stray field) $\mathbf{H}_\mathrm{d}$, governed by the equations
\begin{equation}
\nabla \cdot \mathbf{H}_\mathrm{d} = -\nabla \cdot \mathbf{M},~~~~\nabla \times \mathbf{H}_\mathrm{d} = 0. 
\end{equation}
When simulating two magnets of which one is moving, the demagnetizing field of the stationary magnet can be interpolated at the location of the moving magnet (Fig.~\ref{FIGInterpolation}) and vice versa. The interpolation of $\mathbf{H}_\mathrm{d}$ between cells can be done in two ways: direct interpolation of the field vectors, or by calculating the magnetic scalar potential $\phi_M$, interpolating it and obtaining the demagnetizing field as the gradient of the potential $\mathbf{H}_\mathrm{d} = -\nabla \phi_M$. Both methods are included in our implementation.

A problem that can arise when interpolating the demagnetizing field is introducing artificial divergence and/or curl into the system. It has been shown that artificial divergences induced by simply interpolating the field vectors componentwise can lead to unphysical behavior in magnetohydrodynamics  \cite{divergencefreeneeded}, and thus to avoid similar problems we try to maintain the demagnetizing field as divergence-free (outside the magnet) and curl-free (everywhere) as possible. In this regard, the scalar potential method is advantageous, since it gives a curl-free field by definition as long as the interpolants are $C^2$ continuous, i.e. continuous up to the second derivative \cite{interpb}. The behavior of divergence and curl during interpolation in both scalar potential method and direct interpolation of the field is examined in Section~\ref{sec:results}.

Simulations can include other field terms that require interpolation as well. When applying an external field $\mathbf{H}_\mathrm{ext}$ that is not uniform in the whole simulation domain, the field can change smoothly between cells and therefore has to be interpolated inside the moving magnet. For the external field, one has to directly interpolate the field vectors between cells when the magnet is moving, since the external field is not necessarily expressable as a gradient of a scalar potential. The same applies to the eddy current field. Fortunately the anisotropy field $\mathbf{H}_\mathrm{anis}$ is localized in a cell and the exchange field $\mathbf{H}_\mathrm{exch}$ is very short-ranged, extending only to the nearest neighbor cells. Thus interpolation is not needed for these interactions.

\section{Implementation}
\label{sec:movImplementation}

The movement and interpolation codes were implemented in the micromagnetic solver \texttt{MuMax}3  \cite{vansteenkiste2014design}, due to its open-source nature and authors' previous experience with the software. \texttt{MuMax}3 has functions for moving the entire simulation domain a cell in a specified direction, useful when studying e.g. the movement of a domain wall in a long nanowire. However, since the whole simulation domain is moved, relative motion of two magnets cannot be simulated. Hence we implemented functions with which one can define the part of the simulation domain as moving (''slider'') and part as staying at rest (''base''). The slider part of the simulation domain can then be moved in the desired direction while the base remains in place.
In our extension, the slider can be allowed to move in selected directions in 3 dimensions, making it possible to study perpendicular movement relative to the base (e.g. in adhesion context)  in addition to parallel movement. The slider can be driven either with constant velocity or with a spring moving at a constant velocity. The extension also supports periodic boundary conditions.

The slider and base geometries are defined at the beginning of the simulation, and are assumed not to change. The slider moves as a rigid body, and the location relative to the cell centers (where the exact $\mathbf{m}$ is calculated) is updated constantly, the relevant fields being interpolated according to the location. We incorporated Newton's equations of motion into the Euler solver and the adaptive step RK45 Dormand-Prince solver, solving the equations simultaneously with the LLG equation. Whenever the slider has moved a full simulation cell, the geometry and parameters in the involved cells are updated accordingly and the parameter determining location relative to the cell centers starts again from zero. We modified the calculations of energy and other quantities to use the interpolated fields so that they take into account the partial movement of the slider. Additionally, we introduced new quantities relevant to the motion such as the total forces affecting the base and the slider, the speed and acceleration of the slider etc.

For simulating magnetic friction, we implemented the calculation of some additional measures, such as the force the base exerts on the slider and the power dissipated by the relaxation of $\mathbf{M}$ due to Gilbert damping. Since the slider and base interact only via the demagnetizing field, the force the slider feels from the base can be written as
\begin{equation}
\mathbf{F}_\mathrm{m}^\mathrm{s} = \mu_0V_\mathrm{cell}\sum_{i\in{s}} (\mathbf{M}^\mathrm{s}(\mathbf{x}_i)\cdot \nabla)\mathbf{H}_\mathrm{d}^\mathrm{b}(\mathbf{x}_i),
\label{EQForce}
\end{equation}
where the sum is over the cells designated as the slider, $\mu_0$ is the permeability of vacuum, $V_\mathrm{cell}$ is the volume of a simulation cell, $\mathbf{x}_i$ is the (possibly interpolated) location in \mbox{cell $i$}, and superscripts $s$ and $b$ are used to denote the slider and base, respectively \cite{motioncontrol}.

Moving the slider forward pumps energy into the system with power $P_\mathrm{in} = -\mathbf{F}_\mathrm{m}^\mathrm{s}\cdot \mathbf{v}_\mathrm{s}$, where $\mathbf{v}_\mathrm{s}$ denotes the velocity of the slider. In the steady state, the power pumped into the system and the power dissipated by the relaxation of the magnetic moments (and possible mechanical damping of the spring in the case of spring driving) have to be equal. Power dissipated by the relaxation of $\mathbf{m}$ can be calculated as (\mbox{Ref.~\cite{magiera1}})
\begin{equation}
P_\mathrm{diss} = \frac{\gamma \alpha\mu_0V_\mathrm{cell}}{(1+\alpha^2)M_\mathrm{sat}}\sum_{i=1}^N\big(\mathbf{M}(\mathbf{x}_i)\times \mathbf{H}_\mathrm{eff}(\mathbf{x}_i)\big)^2.
\end{equation}
If other damping factors are small, the friction force determined by energy dissipation $F = \langle P_\mathrm{diss}/ v_\mathrm{s}\rangle$ should then coincide with the force of Eq.~(\ref{EQForce}) in the steady state.

\subsection{Interpolation of the demagnetizing field}
\label{sec:demagImplementation}

A typical way of solving the demagnetizing field in finite difference micromagnetics, including \texttt{MuMax}3, is as a discrete convolution with kernel usually referred to as the demagnetization tensor $\mathbf{N}$. Mathematically the direct calculation of $\mathbf{H}_\mathrm{d}$ at point $\mathbf{r}$ can be written as a convolution integral,
\begin{equation}
\begin{aligned}
\mathbf{H}_\mathrm{d}(\mathbf{r}) &= -\frac{1}{4\pi}\nabla \int \mathbf{M}(\mathbf{r}^\prime)\cdot \nabla^\prime \frac{1}{|\mathbf{r}-\mathbf{r^\prime}|} d^3 \mathbf{r^\prime}\\
&= -\int \mathbf{N}(\mathbf{r}-\mathbf{r}^\prime) \mathbf{M}(\mathbf{r}^\prime) d^3 \mathbf{r^\prime},
\end{aligned}
\end{equation}
of which the discretized version used in the finite difference method is
\begin{equation}
\mathbf{H}_\mathrm{d}(\mathbf{r}_\mathrm{i}) = -\sum_\mathrm{j} \mathbf{N}(\mathbf{r}_\mathrm{i}-\mathbf{r}_\mathrm{j}) \mathbf{M}(\mathbf{r}_\mathrm{j}),
\end{equation}
where $i$ and $j$ denote indices of the discretization cells. The demagnetization tensor is a $3\times3$ matrix containing geometrical coefficients for each pair $\mathbf{r}_\mathrm{i},\mathbf{r}_\mathrm{j}$. The calculation can be sped up by FFTs, changing the convolution in real space to a pointwise multiplication of FFT'd magnetization and the demagnetization tensor \cite{numericalmicromagnetics}, reducing the calculation complexity from $O(n^2)$ to $O(n\log n)$ of FFTs. In addition to performing FFTs, \texttt{MuMax}3 further speeds up the calculations by using the GPU to massively parallelize the effective field calculations. The demagnetization tensor is only computed once, Fourier transformed and moved to the GPU at the beginning of the simulation, while the magnetization has to be transformed and inverse-transformed during each time step, the number of transforms being dependent on the time integration scheme used. 

As for the interpolation of the field values between cells, GPUs can utilize texture memory for fast interpolation. However, these interpolations are linear and the discontinuity of the derivatives make for a poor interpolation for the field. Fortunately, there exists a fast GPU-based interpolation library capable of constructing a 3D cubic b-spline interpolants and finding the gradient at interpolated points, written by Ruijters and Thevenaz \cite{ruijters1}. They have also shown that since b-splines are not truly interpolating functions (meaning the interpolated values differ from the actual point values) by default, the data has to be prefiltered for accurate interpolation \cite{ruijters2}. A function performing the prefiltering is included in the library. We incorporated the library into the extension, and utilize it for both direct componentwise interpolation of the field vectors and interpolation of the scalar potential.

For calculating the demagnetizing field via the scalar potential, Abert \emph{et al.} have presented a method which utilizes a similar kernel multiplication in Fourier space as is done with the demagnetization tensor, yielding a fast way to calculate the scalar potential  \cite{Scalarpotential}. In the scalar potential method, the equation for the demagnetizing field reads
\begin{equation}
\mathbf{H}_\mathrm{d}(\mathbf{r}_\mathrm{i}) = -\nabla\phi_M = -\nabla \sum_\mathrm{j} \mathbf{S}(\mathbf{r}_\mathrm{i}-\mathbf{r}_\mathrm{j}) \mathbf{M}(\mathbf{r}_\mathrm{j}),
\end{equation}
where $\mathbf{S}$ denotes the scalar potential kernel, its elements defined by
\begin{equation}
\mathbf{S}(\mathbf{r}_\mathrm{i}-\mathbf{r}_\mathrm{j}) = \frac{1}{4\pi}\int_{V_j}\nabla^\prime \frac{1}{|\mathbf{r}-\mathbf{r}^\prime|}d^3\mathbf{r}^\prime\bigg\rvert_{\mathbf{r} = \mathbf{r}_\mathrm{i}}.
\end{equation}
We follow their method with a slight modification: instead of calculating the potential in the corners of the cells, we calculate it at the cell centers, since the same kernel can also be used to calculate the solenoidal component of the electric field when simulating eddy currents. The gradient of the scalar potential in our case is obtained by a four-point central finite difference approximation. The demagnetization tensor method gives more accurate results for $\mathbf{H}_\mathrm{d}$ at the boundaries of the magnet, where the field component perpendicular to the boundary can be discontinuous. Thus when using the scalar potential method, we opt to use the scalar potential for the far field only, and use the demagnetization tensor to calculate the near field (the defined magnetization geometry and one additional cell in each direction) since the near field does not require interpolation for either magnet. 

To retain the speed of the simulation at levels comparable to the demagnetization tensor method, we parallelized the scalar potential calculation and gradient on the GPU. Similarly to the demagnetization tensor, the scalar potential kernel is computed and Fourier transformed only once in the beginning of the simulation, and the convolution with the kernel is parallelized easily as a pointwise multiplication. The calculation of the gradient of the potential was also trivially parallelizable.  

We use scalar potential that extends four cells over of the simulation domain boundary in each direction in order to properly prefilter and interpolate the values at the boundaries of the domain. These additional values do not require a larger kernel however, since when not using periodic boundary conditions we get the potential twice the system size in each direction anyway due to the zero padding required for the FFT and kernel multiplication. When using periodic boundaries, the four additional values are simply copied from the other side of the simulation domain.

\subsection{Eddy currents}
\label{sec:EddyImplementation}

If the moving magnets are conducting, the change in magnetic field inside the magnets creates eddy currents, which can impede the motion due to the currents inducing a magnetic field resisting the motion according to Lenz's Law, turning kinetic energy into heat in the process. On the macroscale, this kind of ''eddy current friction'' is utilized in applications such as eddy current brakes \cite{eddycurrentbrakes}. In micromagnetics, eddy currents are usually assumed to be incorporated into the Gilbert damping parameter. When simulated explicitly, they have been found to have a visible effect in the switching times of magnetic nanocubes \cite{TorresEddy,eddyswitching2} and the field strength in magnetic recording heads \cite{eddycurrenthead}. In order to see if and how eddy currents influence the magnetic friction forces, we include an implementation of eddy current simulation in our extension.

Our eddy current implementation follows the method of Torres \emph{et al}, in which the irrotational and solenoidal electric field components, $\mathbf{E}_\mathrm{irrot}$ and $\mathbf{E}_\mathrm{sol}$, are solved separately and summed to obtain the total electric field $\mathbf{E}$. From the electric field, the current density $J = \sigma \mathbf{E}$ and the resulting magnetic field $\mathbf{H}_\mathrm{eddy}$ are then calculated. A short summary of the method is detailed below, for more complete description see Ref.  \cite{TorresEddy}.

The solenoidal electric field is found via the formula
\begin{equation}
\mathbf{E}_\mathrm{sol}(\mathbf{r}_\mathrm{i}) = \sum_\mathrm{j} -\frac{1}{4\pi}\frac{\partial \mathbf{B}(\mathbf{r}_\mathrm{j})}{\partial t} \times \int_{V_j} \frac{\mathbf{r}_\mathrm{i}-\mathbf{r}_\mathrm{j}}{|\mathbf{r}_\mathrm{i}-\mathbf{r}_\mathrm{j}|^3} d^3\mathbf{r}_\mathrm{j},
\end{equation}
where $\mathbf{B} = \mu_0(\mathbf{H}_\mathrm{d} + \mathbf{H}_\mathrm{ext} + \mathbf{M})$. Noting that 
\begin{equation}
\int_{V} \frac{\mathbf{r}-\mathbf{r}^\prime}{|\mathbf{r}-\mathbf{r}^\prime|^3} d^3\mathbf{r}^\prime = \int_{V}\nabla^\prime \frac{1}{|\mathbf{r}-\mathbf{r}^\prime|}d^3\mathbf{r}^\prime,
\end{equation}
we see that the scalar potential kernel $\mathbf{S}$ can also be utilized in the calculation the solenoidal electric field,
\begin{equation}
\mathbf{E}_\mathrm{sol}(\mathbf{r}_\mathrm{i}) = \sum_\mathrm{j}\frac{\partial \mathbf{B}(\mathbf{r}_\mathrm{j})}{\partial t} \times \mathbf{S}(\mathbf{r}_\mathrm{i}-\mathbf{r}_\mathrm{j}).
\end{equation}
Taking the cross product of the scalar potential kernel and Fourier transformed $\partial \mathbf{B}(\mathbf{r}_\mathrm{j})/{\partial t}$ on the GPU, we find the solenoidal electric field. 

Assuming charge neutral material, the irrotational field $\mathbf{E}_\mathrm{irrot}$ can be calculated from the electric scalar potential $\phi_E$, which is found by solving the Laplace's equation
\begin{equation}
\Delta \phi_E = 0
\end{equation}
inside the magnet, with the boundary condition $\partial \phi_E / \partial n = \mathbf{E}_{\mathrm{sol},n}$, where $n$ denotes the surface normal of the magnet. This boundary condition ensures that the eddy currents are tangential to the surface. In our implementation, the Laplace's equation for the electric potential is solved iteratively via successive-over-relaxation performed on the GPU. Taking the gradient of the potential then yields the irrotational electric field.

The current density $\mathbf{J}$ is obtained by summing the electric fields $\mathbf{E}_\mathrm{irrot}$ and $\mathbf{E}_\mathrm{sol}$ and multiplying with conductivity $\sigma$, which is treated as a uniform constant across the material. From the current density, another tensor multiplication with the scalar potential kernel, computed in the same fashion as the solenoidal electric field, is required to obtain the magnetic field $\mathbf{H}_\mathrm{eddy}$ generated by eddy currents.

\section{Numerical examples}
\label{sec:results}

\subsection{Divergence and curl}

The effect of interpolation on the divergence and curl of the demagnetizing field in both scalar potential method and direct componentwise interpolation of $\mathbf{H}_\mathrm{d}$ was studied by creating a completely random magnetization in the base film and keeping the magnetic vectors frozen during movement. The demagnetizing field of the base was calculated and interpolated inside the slider while observing how the interpolation affects the average magnitudes of divergence and curl, defined by
\begin{equation}
\begin{aligned}
D(x) &= \frac{1}{N}\sum_i^N|\nabla \cdot \mathbf{H}_\mathrm{d}|_i(x)\\
C(x) &=\frac{1}{N}\sum_i^N||\nabla \times \mathbf{H}_\mathrm{d}||_i(x),
\end{aligned}
\end{equation}
where the divergences and curls were calculated in each cell $i$ using a two-point central finite difference approximation, and $x$ refers to the partial movement between cells. Since the method and numerical noise always give some nonvanishing divergence and curl, we use the divergence and curl of the stationary non-interpolated field as a baseline and calculate the relative difference of the interpolated values compared to the stationary value. 

\begin{figure}[t!]
\leavevmode
\includegraphics[trim=0cm 0cm 0cm 0cm, clip=true,width=1.0\columnwidth]{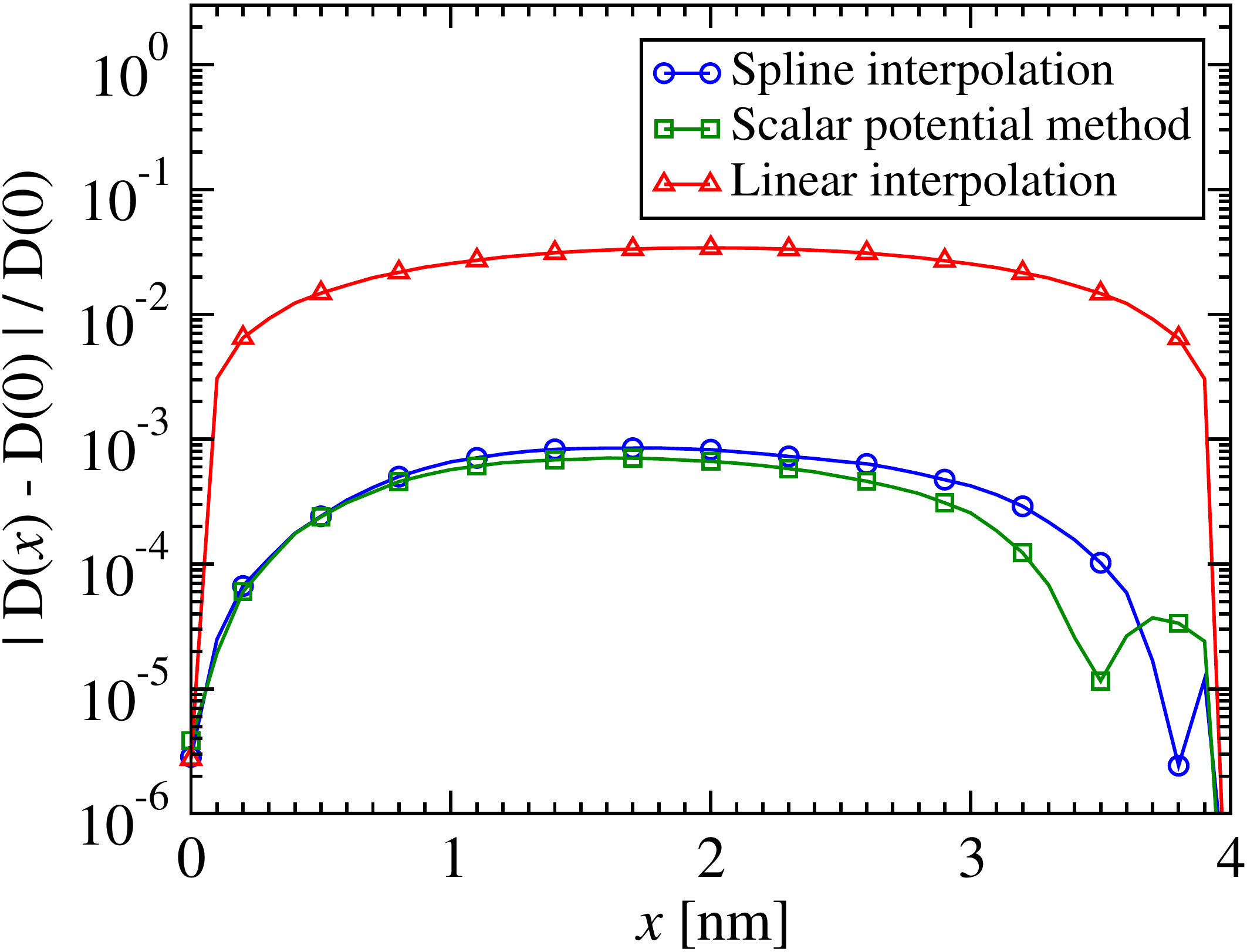}
\caption{The relative error of divergence during the interpolation between two cell centers 4 nm apart, $x$ denoting the location of the slider. The spline interpolation is slightly asymmetric, the error reversing sign close to the end of the interpolation.}
\label{FIGDiviCurli}
\end{figure}

The results for divergence are gathered in Fig.~\ref{FIGDiviCurli}. As can be seen from the figure, the interpolation induces relatively small errors to the divergence of the field during the interpolation in both the scalar potential method and direct interpolation, even in the case of a randomized magnetization resulting in a mostly random field. In the case of a smoother field, the errors are likely to be smaller. For reference, the effect of linearly interpolating the field vectors was also studied, and though one or two orders of magnitude greater difference to the stationary case than with the spline interpolation, even in the case of linear interpolation the difference is in the order of percents. The results are similar for the curl of the field.

\subsection{Magnetic friction}
For simulating magnetic friction, we used a system of two \mbox{1024 nm $\times$ 1024 nm $\times$ 20 nm} thin films with a distance of 20 nm and periodic boundary conditions in the film plane. The domain discretized in cubic cells with 4 nm side length. We used CoCrPt as the material, with parameters similar to those in Ref. \cite{navas2014domain}, i.e. uniaxial anisotropy in the $+z$-direction, and \mbox{$K_\mathrm{u} = 1.225 \cdot 10^5$ J/m$^3$}, \mbox{$M_\mathrm{sat} = 3.5\cdot 10^5$ A/m}, \mbox{$A_\mathrm{ex} = 5\cdot 10^{-12}$ J/m} and \mbox{$\alpha = 0.05$}. Disorder was introduced by dividing the upper and lower films into grains of 20 nm average size using Voronoi tessellation \cite{disorder}, and setting the direction of the anisotropy vector randomly from 0$^\circ$ to $8^\circ$ from the $+z$-axis for each grain. The films were initialized to a stripe pattern similar to Ref.~\cite{motioncontrol}, with approximately 80 nm wide stripes and let relax (Resulting in a system similar to what was shown in Fig.~\ref{FIGSimulationExample}), after which the upper film was driven in the $+x$-direction by a spring moving at a constant velocity $v_\mathrm{d} = 2$ m/s for 300 ns. The friction force was measured from the spring elongation $F_\mathrm{spring}=k(v_\mathrm{d}t - x_\mathrm{s})$, where $x_\mathrm{s}$ is the position of the slider. The spring constant was chosen as \mbox{$k = 0.005$ N/m}.  The simulations were carried out in zero temperature. We also included a viscous damping term in the equation of motion, 
\begin{equation}
\mathbf{F}_\mathrm{d} = -\gamma m \dot{x}_\mathrm{s},
\end{equation}
where $m$ is the mass of the slider, $\dot{x}_\mathrm{s}$ its velocity and $\gamma$ is a viscous damping coefficient. The complete equation of motion for the slider is then 
\begin{equation}
m\ddot{x}_\mathrm{s} = k(v_\mathrm{d}t - x_\mathrm{s}) - \gamma m \dot{x}_\mathrm{s} + F^\mathrm{s}_\mathrm{m},
\end{equation}
where $F^\mathrm{s}_\mathrm{m}$ is the $x$-directional component of the force exerted by the base on the slider defined in Eq.~(\ref{EQForce}). The mass and the damping coefficient in these numerical examples are chosen such that the spring-slider system is critically damped and the resulting viscous damping force is roughly an order of magnitude smaller than $F^\mathrm{s}_\mathrm{m}$ when $v_\mathrm{s} = v_\mathrm{d}$. To see whether the film distance would affect the friction behavior in our tests as it did in Ref. \cite{motioncontrol}, we ran simulations for various film distances, ranging from 20 nm to 80 nm. We started the simulation with the spring already ahead of the slider by 80 nm so that the beginning part of the simulation, where the spring slowly elongates increasing the force, is shorter.

\begin{figure}[t!]
\leavevmode
\includegraphics[trim=0cm 0cm 0cm 0cm, clip=true,width=1.0\columnwidth]{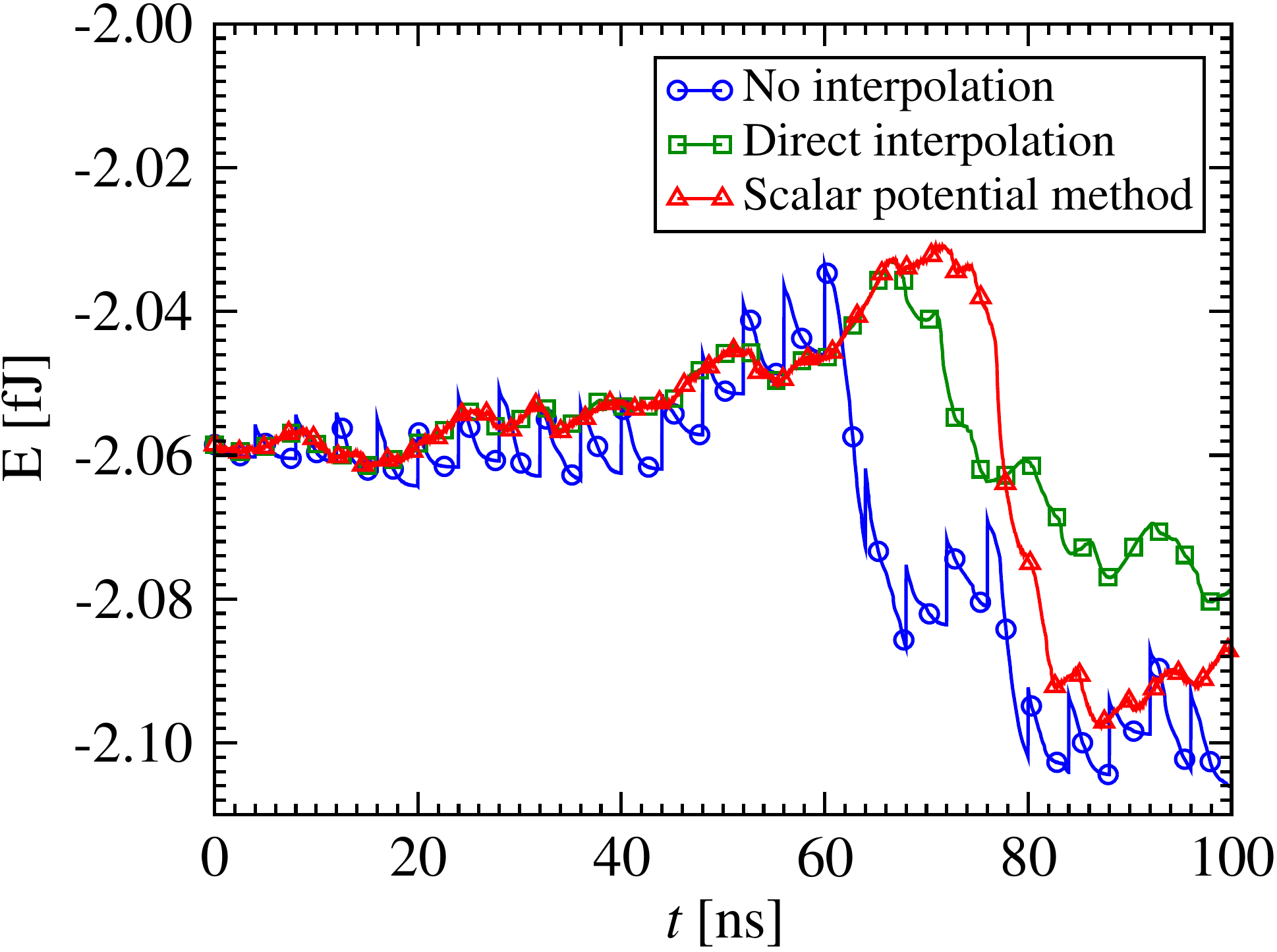}
\caption{Energies for different interpolation methods. Though the dissimilar time evolution of the magnetization is reflected by the energy of the system in the various cases, the trend in the total energy is similar in all three cases. The drop in energy is the result of a magnetization reconfiguration in response to motion of the slider.}
\label{FIGEnergyinterpolation}
\end{figure}

A snapshot of the system total energy for the different interpolation methods in an example simulation can be seen in Fig.~\ref{FIGEnergyinterpolation}. For both the direct componentwise interpolation and the scalar potential interpolation, the system becomes continuously driven instead of the periodic jumps of the non-interpolated case and thus the discontinuities in the total energy are eliminated, as expected.

For a relatively long simulation time, both the direct interpolation and the scalar potential method predict the same magnetization dynamics and hence the same energy. When driven for long enough, small discrepancies in the predicted field values can lead to slightly different time evolution of domains. However, the effect this has on quantities such as the average friction force is minimal.

\begin{figure}[t!]
\leavevmode
\includegraphics[trim=0cm 0cm 0cm 0cm, clip=true,width=1.0\columnwidth]{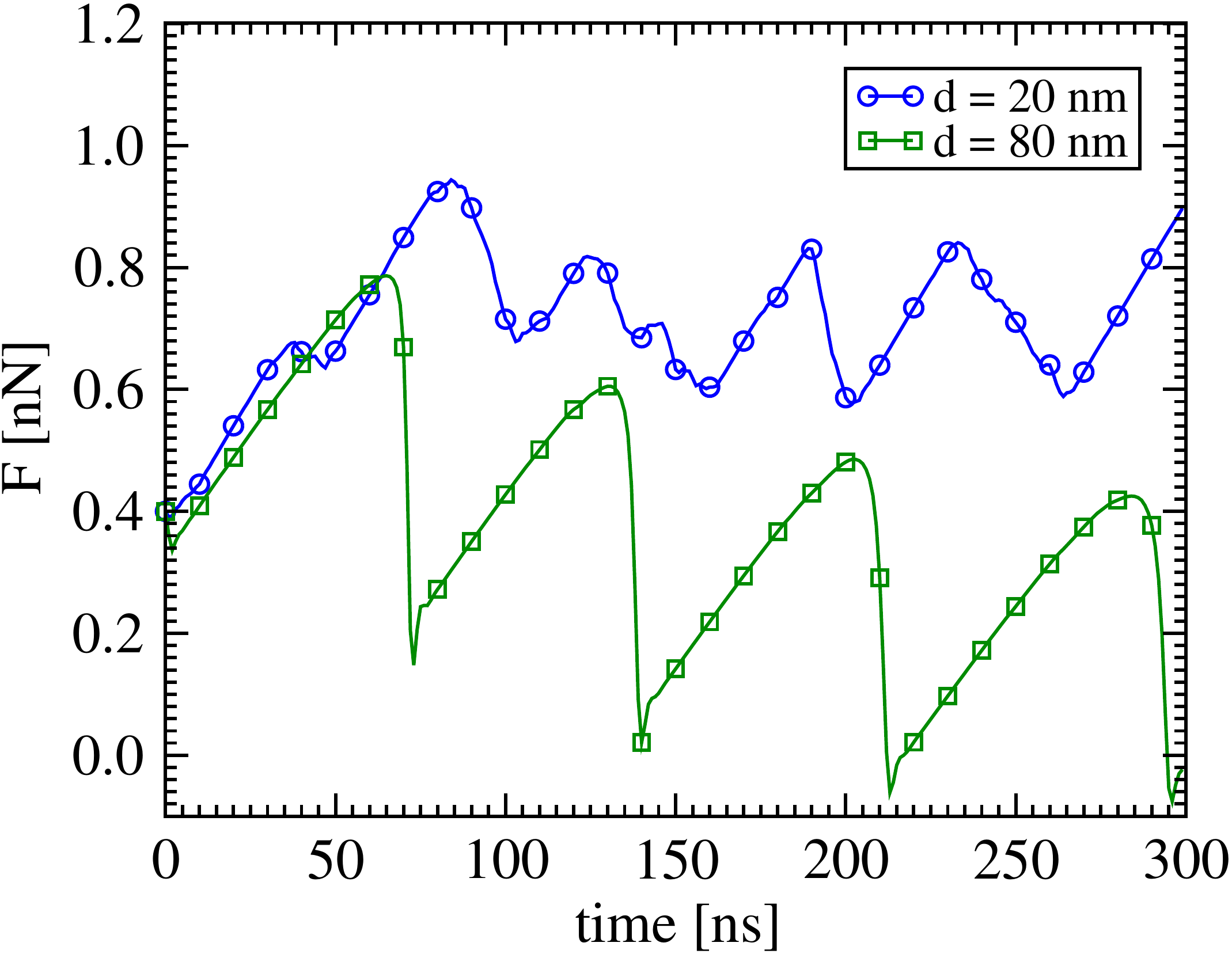}
\caption{The friction forces with inter-film distance of \mbox{20 nm}, in which the stripe domains in the slider deform but are held in place by the stripes of the base, and \mbox{80 nm}, where the slider experiences stick-slip motion, and the stripes experience only minor changes. The diminishing force in the \mbox{80 nm} case is due to the stripes having slightly different widths, and the initial configuration having better matching stripes in the slider and base.}
\label{FIGFrictionForce}
\end{figure}
Considering the fact that we used different material and way of realizing disorder along with smaller films, the qualitative behavior and the numerical results are comparable to those of Ref.~\cite{motioncontrol}. With little distance between the magnetic films, the interaction between the stripe domains of the base and slider are stronger than the pinning effect of the disorder in the slider, and thus the stripes are locked in place. Due to the movement of the slider, the stripe domains deform, but they don't move with the slider and thus the slider is actually pulled through its own stripe domains. In this case, the magnetic domain configuration changes constantly during the dragging, and thus the energy dissipation due to relaxation of magnetic moments is large compared to the damping of the spring. In this case the average friction force calculated from power dissipation and directly from the force exerted by the base on the slider are roughly equal, giving approximately 0.7 nN as the friction force. 

An increase in the distance leads to the stripe domains getting pinned in place inside the slider by the disorder. This results in a situation where the slider and it's pinned stripe domains stick and slip in the periodic potential created by the domain structure of the base, alternating between increase of the force during stick phase and rapid decrease during slips (Fig.~\ref{FIGFrictionForce}). Even though the stripe domains provide resistance to the motion in the form of potential wells, the stripes themselves deform comparatively little, and thus the energy dissipation is actually dominated by the damping of the spring, which grows comparatively large during the slips. The contribution of the magnetic moment relaxation to the average friction force was only \mbox{0.04 nN} in this case. 

In our simulations, even with the smallest distance between films there was some behavior reminiscent to stick-slip in addition to domain dragging. This resulted from a stripe sticking to individual grains with strong anisotropy before ''snapping'' to another configuration. The stripe domains realigning with the driving direction was also observed in some configurations similar to Ref.~\cite{motioncontrol}. 

\subsection{Eddy current simulations}

The eddy current implementation was tested by running a simulation in which a Permalloy nanocube switches its magnetization due to an external field, similar to the example simulation of Ref.~\cite{TorresEddy}. The Permalloy cube with side length of 40 nm was discretized into cubic cells of 2.5 nm side length. The magnetization was first relaxed into a $+z$-directed flower state \cite{flowerstate}, and the switching was caused by an external magnetic field in the $-z$ direction. The field strength was ramped linearly from zero to \mbox{$100$ mT} in the span \mbox{of 0.1 nanoseconds}, after which the field is held constant. The simulations were run in zero temperature, with \mbox{$A_{ex} = 13\cdot 10^{-12}$ J/m}, \mbox{$M_{sat} = 860\cdot 10^3$ A/m} and varying \mbox{$\alpha = 0.01 - 0.1$}. The value used for conductivity of Permalloy was \mbox{$\sigma = 6.25\cdot 10^6$ S/m}. \cite{torres2}

\begin{figure}[t!]
\leavevmode
\includegraphics[trim=0cm 0cm 0cm 0cm, clip=true,width=1.0\columnwidth]{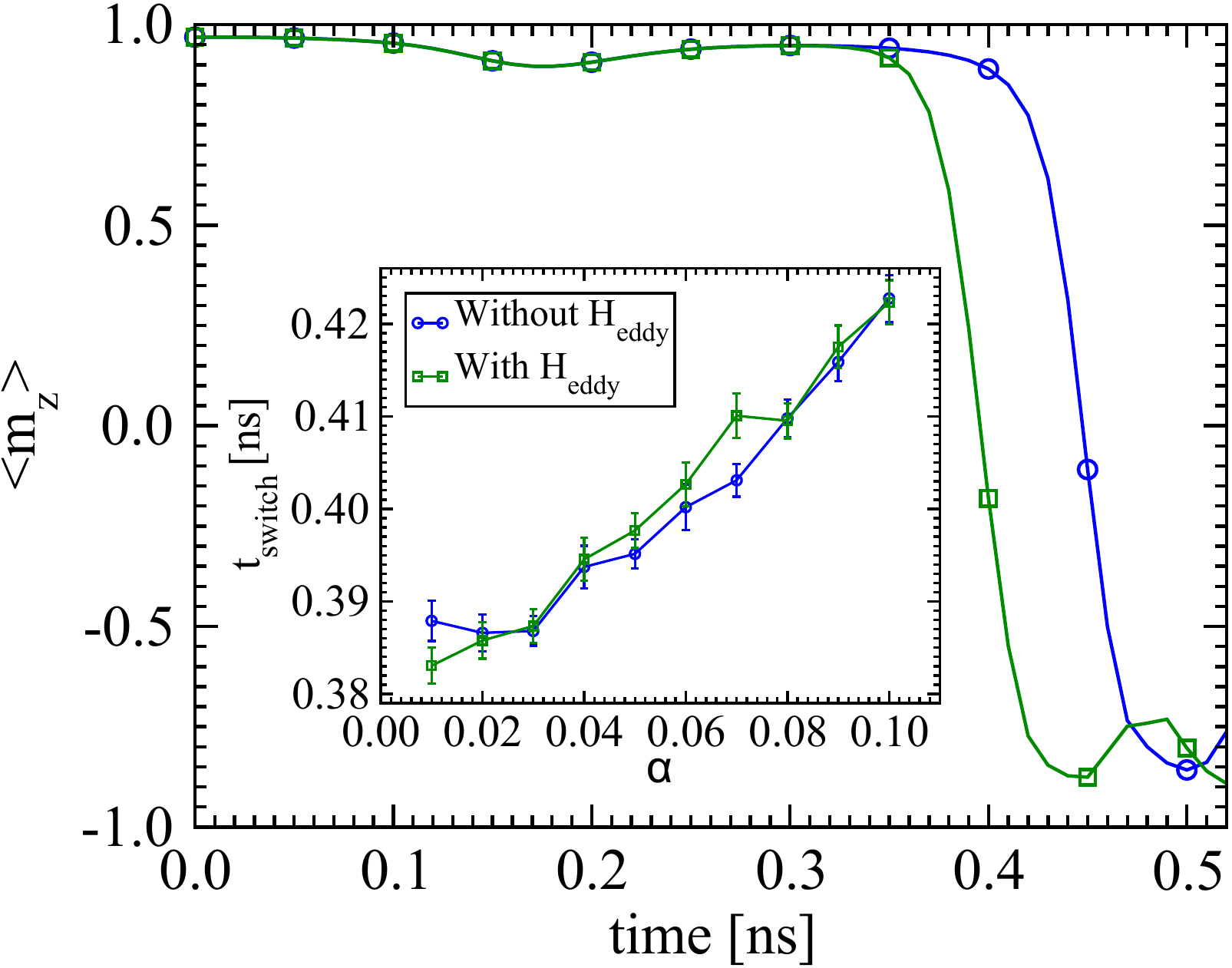}
\caption{The anticipation of the magnetic switching of a Permalloy nanocube with $\alpha = 0.01$ when eddy currents are included in the simulation compared to the simulation without eddy currents. The inset shows that on average, the eddy currents anticipate the switching with low values for $\alpha$, while the trend actually seems to reverse for intermediate values. For the highest three values the difference in average switching time is negligible.}
\label{FIGeddya005}
\end{figure}

An example case for $\alpha = 0.01$ is shown in Fig.~\ref{FIGeddya005}, and in this simulation the switch is anticipated by the eddy currents similarly as in previous literature  \cite{TorresEddy,eddyswitching2}. However, we found the simulation to be sensitive to the small changes in the initial conditions. For example, relaxing the magnetization for a few more picoseconds could in some cases influence the switching time by almost a hundred picoseconds even without eddy currents. The sensitivity might be due to the switching being an avalanche-like event, starting from small rotations in the magnetization vectors which generate the eddy field that in turn affects the time-evolution of the magnetization. Thus small changes in the eddy field and/or magnetization can have relatively large effects on the final switching time. 

To mitigate the sensitivity to initial conditions, we averaged the simulations over 150 realizations, each relaxed and then slightly perturbed with a weak random field. Though most of the results for averaged behaviour with and without eddy currents (inset of Fig.~\ref{FIGeddya005}) are within the standard error of the mean from each other, the results suggest that the effect of eddy currents depends on the value of the Gilbert damping. This seems reasonable, since after the initial contribution of the change in external magnetic field, the only contribution to eddy currents comes from the change in magnetization $\partial\mathbf{M}/\partial t$, which depends on the damping parameter. As noted in \mbox{Ref.~\cite{eddycurrenthead}}, eddy currents provide additional resistance to the precessional motion of the magnetic moments. In the low $\alpha$ regime, the magnetization tilts initially towards the $xy-$plane, after which there is a brief out-of-plane rotation, and then the switching occurs. The lessened precession with eddy currents can lead to magnetization not tilting as much out of the $xy$-plane after the initial tilt, thus leaving the magnetization more susceptible to initiating the switch. For \mbox{$\alpha =$ 0.05 - 0.07}, the additional resistance to precession due to eddy currents appears to act similarly to a higher damping constant, delaying the switching. For the largest values of \mbox{$\alpha =$ 0.08 - 0.10}, the switching is likely dominated overall by the Gilbert damping and thus the effect of eddy currents is negligible.

The effect of eddy currents on the magnetic friction force was also studied, with the same simulation scenario of two CoCrPt films as in the movement simulations before, but with eddy currents included. In this test, we used smaller films (512 nm $\times$ 512 nm in the $xy$-plane) for shorter computation times. Not finding a documented value of the conductivity of thin CoCrPt films, we chose a conductivity similar to Permalloy, $\sigma = 10^6$ S/m. 

When compared to the movement simulation without eddy currents, there's no difference in the force resisting the motion at the first 100 ns. Later, the force changes compared to the eddy currentless case, though in our test simulation, the difference is relatively small (few tens \mbox{of pN}). The change seems to be caused less by the eddy current field of the base directly affecting the slider (as $\mathbf{H}_\mathrm{eddy}$ at that distance is very weak) and more by the change in the magnetization in both the base and the slider due to the eddy current fields inside the films. This change in magnetization further affects the $\mathbf{H}_\mathrm{exch}$ and the $\mathbf{H}_\mathrm{d}$, modifying the time evolution of the magnetization and thus the perceived force. In order to gain a clearer picture about how eddy currents affect the movement, a more in-depth study is required. Additionally, since Joule heating plays a significant role in the energy dissipation due to eddy currents, it would likely have to be taken into account in order to obtain conclusive results.

\subsection{Performance and limitations}
The scalar potential method requires fewer FFTs and calculations in the reciprocal space, and this somewhat mitigates the simulation time increase brought by the interpolation and the calculation of near field via the demagnetization tensor. Since the scalars also require fewer interpolations overall than do vectors, both the scalar potential method and the direct interpolation via splines are quite equal in computational cost. The discrete movement is naturally the fastest, as no additional calculations are required compared to simulations without movement. However, since one has to calculate the force exerted by the demagnetizing field of the base on the slider film, one still has to calculate the fields separately. This along with some calculations related to the movement results in the modified RK45 solver being a bit slower than the RK45 solver of \texttt{Mumax}3 in general. 

Eddy currents are computationally quite heavy, requiring two extra FFTs, IFFTs and interpolations for both the base and the slider, and the solving of a single Laplace's equation. Depending on the desired accuracy, they increase the computation time from roughly double up to an order of magnitude. The calculation of eddy currents also requires the information on the surface normals, and as such the simulated magnets have to be more than 1 cell thick or the normal is not uniquely defined.  Additionally, since small enough changes in magnetization induce an eddy field so weak that it can be lost due to floating point precision, the eddy currents might start to affect the magnetization evolution at different points of time with different parameters. 

Since some quantities are calculated using a two-point central finite difference approximation, the values of the demagnetizing field in cells just outside the boundary of the base and slider in each direction are also interpolated to give correct values inside the magnet. Hence the slider and base have to be more than 2 simulation cells apart, which typically means distances larger than approx. \mbox{5 - 10 nm} depending on cell size. 

We found that when doing the spline interpolation on the GPU, performing 64 lookups on the nearest neighbors instead of 8 trilinear interpolations for the construction of the spline gives more accurate results and less noise. This might be caused by the fact that the linear interpolations on the GPU can take only 254 possible coordinate positions between two texture points \cite{ruijters1}, and thus using trilinear interpolations we implicitly round the position between cells to one of these 254 values, whereas with the 64 nearest neighbor lookups we get the position to floating point accuracy.\\

\section{Summary}
\label{sec:summary}

We have augmented an existing micromagnetic code \texttt{Mumax}3 with the possibility of moving geometries independently inside the simulation domain, making it possible to study two magnets in relative motion in a micromagnetic framework. We implemented smooth motion through interpolating the fields affecting the magnets during movement. For the interpolation of field terms in the LLG equation, an external library for 3D cubic spline interpolation on the GPU was integrated to the code, as well as a method for calculating the demagnetizing field with the scalar potential. The extension source code has been published as a separate branch for \texttt{Mumax}3 on GitHub \cite{github}.

We tested our movement and eddy current implementations with various numerical example simulations and demonstrated that smooth relative motion can be well approximated using splines to interpolate the demagnetizing fields, while the errors in divergence and curl of the fields remain minimal. Comparison with the results from previous literature indicates that finite difference micromagnetic simulations are a suitable framework for studying the motion of microscale magnets and phenomena related to the motion such as magnetic friction.

Eddy currents were found to have an effect in the switching time of Permalloy nanocubes and the friction force between two thin films in relative motion. However, the exact mechanism of how eddy currents and the related parameters affect the switching time merit further study. Additionally, more simulations are required to better assess the effect of eddy currents in magnetic friction context.

\begin{acknowledgments}
We acknowledge the support of the Academy of 
Finland via an Academy Research Fellowship (LL, projects no. 268302 and 303749), and 
the Centres of Excellence Programme (2012-2017, project no. 251748). We acknowledge 
the computational resources provided by the Aalto University School of Science 
Science-IT project, as well as those provided ny CSC (Finland). 
\end{acknowledgments}

\bibliography{bibl}

\end{document}